\documentclass[12pt,letterpaper]{article}
\usepackage[pdftex]{graphicx,color}
\usepackage{hyperref}
\usepackage{amsmath,amssymb}
\usepackage[dvips,letterpaper,text={6.5in,9in}]{geometry}
\usepackage{fancyhdr}
\usepackage{verbatim}

\newcommand\ltap{\
  \raise.3ex\hbox{$<$\kern-.75em\lower1ex\hbox{$\sim$}}\ }
\newcommand\gtap{\
  \raise.3ex\hbox{$>$\kern-.75em\lower1ex\hbox{$\sim$}}\ }

\newcommand\simge{\mathrel{%
   \rlap{\raise 0.511ex \hbox{$>$}}{\lower 0.511ex \hbox{$\sim$}}}}
\newcommand\simle{\mathrel{
   \rlap{\raise 0.511ex \hbox{$<$}}{\lower 0.511ex \hbox{$\sim$}}}}

\newcommand{\slashchar}[1]%
        {\kern .25em\raise.18ex\hbox{$/$}\kern-.75em #1}
\def\lsim{\mathrel{\raise.3ex\hbox{$<$\kern-.75em\lower1ex\hbox{$\sim$}}}}
\def\gsim{\mathrel{\raise.3ex\hbox{$>$\kern-.75em\lower1ex\hbox{$\sim$}}}}

\newcommand\CB{{\cal B}}

\newcommand\CH{{\cal H}}

\newcommand\CO{{\cal O}}

\newcommand\be{\begin{equation}}
\newcommand\ee{\end{equation}}
\newcommand\bea{\begin{eqnarray}}
\newcommand\eea{\end{eqnarray}}
\newcommand\ba{\begin{array}}
\newcommand\ea{\end{array}}

\newcommand\gev{{\rm GeV}}

\newcommand\ifb{{\rm fb}^{-1}}

\newcommand\ellm{\ell^-}

\newcommand\ellp{\ell^+}

\begin{document}

\title{
\vskip -15mm
\begin{flushright}
 \vskip -15mm
 {\small CERN-PH-TH-2014-229\\
   LAPTH-227/14\\
 }
 \vskip 5mm
 \end{flushright}
{\Large{\bf Lepton Flavor Violation in $B$ Decays?}}\\
} \author{
  {\large Sheldon L.~Glashow$^{1}$\thanks{slg@physics.bu.edu} ,\,
  Diego Guadagnoli$^{2}$\thanks{diego.guadagnoli@lapth.cnrs.fr}}\, and 
  Kenneth Lane$^{1}$\thanks{lane@physics.bu.edu}\\
{\large $^{1}$Department of Physics, Boston University}\\
{\large 590 Commonwealth Avenue, Boston, Massachusetts 02215}\\
{\large $^{2}$Laboratoire d'Annecy-le-Vieux de Physique Th\'eorique} \\
{\large UMR5108\,, Universit\'e de Savoie, CNRS} \\
{\large B.P.~110, F-74941, Annecy-le-Vieux Cedex, France}\\
}
\maketitle

\begin{abstract}
  
  The LHCb Collaboration's measurement of $R_K = \CB(B^+ \to K^+
  \mu^+\mu^-)/\CB(B^+ \to K^+ e^+e^-)$ lies 2.6$\sigma$ below the Standard
  Model prediction. Several groups suggest this deficit to result from new
  lepton non-universal interactions of muons. But non-universal leptonic
  interactions imply lepton flavor violation in $B$-decays at rates much
  larger than are expected in the Standard Model. A simple model shows that
  these rates could lie just below current limits. An interesting consequence
  of our model, that $\CB(B_s \to \mu^+\mu^-)_{exp}/\CB(B_s \to \mu^+\mu^-)_{SM}
  \cong R_K \cong 0.75$, is compatible with recent measurements of these
  rates. We stress the importance of searches for lepton flavor violations,
  especially for $B \to K\mu e$, $K\mu\tau$ and $B_s \to \mu e$, $\mu\tau$.

\end{abstract}


\newpage


The LHCb Collaboration recently measured the ratio of
decay rates for $B^+ \to K^+ \ellp\ellm$ ($\ell = \mu,e$),
obtaining~\cite{Aaij:2014ora}:
\be\label{eq:RK}
R_K \equiv \frac{\CB(B^+ \to K^+ \mu^+\mu^-)}{\CB(B^+ \to K^+ e^+e^-)} = 
0.745^{+0.090}_{-0.074}\,{\rm (stat)}\pm 0.036\,{\rm (syst)}\,.
\ee
This result is a 2.6$\sigma$ deficit from the standard model (SM) expectation,
$R_K = 1+\CO(10^{-4})$~\cite{Bobeth:2007dw,Bouchard:2013mia,Hiller:2003js}.

Previous measurements of $R_K$ by the Belle and BaBar
Collaborations~\cite{Wei:2009zv,Lees:2012tva} had considerably greater
uncertainties but were consistent with the SM prediction. The LHCb
determination was made for $1 < q^2 = M^2_{\ell\ell} < 6\,\gev^2$ in order to
be well below the radiative tail of the $J/\psi$. LHCb also measured the $B
\to K \mu^+\mu^-$ branching ratio in this $q^2$-range~\cite{Aaij:2012vr}. The
updated result, based on its full Run~I dataset of $3\,\ifb$,
is~\cite{Aaij:2014pli}
\be\label{eq:NewBRKmumu}
\CB(B^+ \to K^+\mu^+\mu^-)_{[1,6]} = (1.19 \pm 0.03 \pm 0.06)\times 10^{-7}\,.
\ee
The SM prediction~\cite{Bobeth:2011gi,Bobeth:2011nj,Bobeth:2012vn}
\be\label{eq:SMBKmumu}
\CB(B^+ \to K^+ \mu^+ \mu^-)^{\rm SM}_{[1,6]} = 
(1.75^{+0.60}_{-0.29}) \times 10^{-7}\,,
\ee
is about $45\%$ higher. Whether this is the cause of the $R_K$ deficit is
suggestive, but not certain. Another possibility is a problem in the
$B \to K e^+ e^-$ measurement, since LHCb performs better in the $\mu\mu$
than in the $ee$ channel, the latter being hampered by bremsstrahlung and poorer
statistics~\cite{Aaij:2014ora}. On the other hand, bremsstrahlung effects largely
cancel in the experimental observable, and results for $B \to K e^+ e^-$ are
consistent with SM expectations.

The LHCb results for the $B \to K^{(*)} \mu^+\mu^-$ angular distributions and
$R_K$ have attracted much theoretical attention~\cite{Descotes-Genon:2013wba,
  Altmannshofer:2013foa,Beaujean:2013soa,Hurth:2013ssa,Altmannshofer:2014cfa,
  Gauld:2013qja,Gauld:2013qba,Datta:2013kja,Buras:2013qja,Buras:2013dea,
  Horgan:2013pva,Buras:2014fpa,Hiller:2014yaa,Crivellin:2014kga,
  Ghosh:2014awa,Biswas:2014gga,Hurth:2014vma}.
References~\cite{Altmannshofer:2013foa,Altmannshofer:2014cfa,
  Gauld:2013qja,Gauld:2013qba,Datta:2013kja,Buras:2013qja,Buras:2013dea,
  Buras:2014fpa,Hiller:2014yaa,Biswas:2014gga} proposed the existence of
particles at or above 1~TeV which induce new and {\em non-universal} lepton
interactions. However, any departure from lepton universality is necessarily
associated with the violation of lepton flavor conservation. {\em No known
  symmetry principle can protect the one in the absence of the other.} Thus,
LHCb's reported value of $R_K$ implies, e.g., that $B \to K^{(*)} \mu^\pm
e^\mp$ and $B \to K^{(*)} \mu^\pm \tau^\mp$ must occur at rates much larger
than would occur in the SM due to tiny neutrino masses. We urge that these
and other lepton flavor violations (LFV) be sought with renewed vigor in LHC
Run~II and elsewhere.

To illustrate the sort of LFV processes that might be seen in $B$ decays, the
limits that currently exist on lepton mixing, and the potential for discovery
of LFV, we consider a simple but well-motivated interaction that that can
account for the known features of the $R_K$ deficit, is consistent with
existing limits on LFV, and produces effects that may be observable in LHC
Run~II. The LHCb data on $B \to K^{(*)} \ellp\ellm$ suggests that, despite
the detector's superior measurement of muons vis-\`a-vis electrons, the $R_K$
result is due to a $\sim 25\%$ deficit in the muon channel. If so, then LFV
is larger for muons than for electrons. This is naturally accounted for by a
third-generation interaction of the type that would be expected, e.g., in
topcolor models~\cite{Hill:1994hp}. Furthermore, recent theoretical
analyses~\cite{Ghosh:2014awa,Descotes-Genon:2013wba,Altmannshofer:2013foa}
indicate that the Hamiltonian for $B \to K^{(*)}\mu^+\mu^-$ (and $B_s \to
\mu^+\mu^-$) is best described by the SM and new physics (NP) terms
\be\label{eq:bsmm}
\CH_{SM+NP}(\bar b \to \bar s\mu^+\mu^-) \cong
-\frac{4G_F}{\sqrt{2}}V_{tb}^* V_{ts}\frac{\alpha_{EM}(m_b)}{4\pi}
\left[\bar b_L \gamma^\lambda s_L \, \bar \mu \left(C_9^\mu \gamma_\lambda 
 + C_{10}^\mu \gamma_\lambda \gamma_5\right)\mu \right] + 
{\rm H.c.}\,,
\ee
where $C_i^\mu = C_{i,SM}^\mu + C_{i,NP}^\mu$ are Wilson coefficients with
$C_9^\mu \cong -C_{10}^\mu > 0$. The NP contributions to these coefficients
are opposite in sign to the SM contributions, i.e., the lepton current is
approximately of $V$$-$$A$ form. Therefore, we assume that the
third-generation interaction giving rise to the NP part of this Hamiltonian
is
\be\label{eq:HNP} 
\CH_{NP} = G\, \bar b'_L\gamma^\lambda b'_L \, \bar\tau'_L \gamma_\lambda
\tau'_L\,, 
\ee
where $G$ is a new-physics Fermi constant ($G = 1/\Lambda_{NP}^2 \ll G_F$)
and the primed fields are the same as appear in the electroweak currents and
the Yukawa couplings to the Higgs boson. This sort of interaction would arise
from heavy $Z'$-exchange in a topcolor model.\footnote{Two other
  possibilities in this vein are the interactions $G\, \bar
  b'_L\gamma^\lambda s'_L \bar \mu'_L \gamma_\lambda \tau'_L + {\rm H.c.}$ or
  $G\, \bar b'_L\gamma^\lambda s'_L \bar \tau'_L \gamma_\lambda \mu'_L + {\rm
    H.c.}$. The first separately conserves the generation numbers $N_2$ and
  $N_3$ while the second conserves $N_2 + N_3$. Also see the references for
  other $Z'$ models of the anomalies in $B \to K^{(*)} \ell^+ \ell^-$} These
fields are related to the mass-eigenstate (unprimed) fields by unitary
matrices $U_L^d$ and $U_L^\ell$:
\be\label{eq:Umatrices}
d'_{L3} \equiv b'_L = \sum_{i=1}^3 U^d_{L3i} \, d_{Li}\,,\quad
\ell'_{L3} \equiv \tau'_L = \sum_{i=1}^3 U^\ell_{L3i}\, \ell_{Li}\,.
\ee
In particular, the NP interaction responsible for the $R_K$ deficit is
\be\label{eq:bsmmNP}
\CH_{NP}(\bar b \to \bar s \mu^+\mu^-) = G\left[U^{d*}_{L33} U^d_{L32}\vert
  U^\ell_{L32}\vert^2 \, \bar b_L \gamma^\lambda s_L \, \bar \mu_L
  \gamma_\lambda \mu_L + {\rm H.c.}\right]\,.
\ee
The hierarchy of the CKM matrix for quarks and the apparent preference of the
new physics for muons over electrons suggest that $|U^{d,\ell}_{L31}|^2 \ll
|U^{d,\ell}_{L32}|^2 \ll (U^{d,\ell}_{L33})^2 \cong 1$. These expectations
can be tested in searches for $B \to K \mu e$ vs.~$K \mu\tau$. Since the
coefficient of the SM term in Eq.~(\ref{eq:bsmm}) is positive, we assume that
$G\,U^d_{L32} < 0$. The reduction of the SM strength in Eq.~(\ref{eq:bsmm})
by this interaction is also supported by the LHCb measurement of the quantity
$P'_5$ in $B^0 \to K^{*0} \mu^+ \mu^-$ angular distributions in the low-$q^2$
region. Integrated over $1.0 < q^2 < 6.0\,\gev^2$, the $P'_5$ deficit
amounts to 2.5$\sigma$~\cite{Aaij:2013qta}.

Up to the matrix elements of $\bar b_L\gamma^\lambda s_L \, \bar\ell_L
\gamma_\lambda \ell_L$, the $B \to K^{(*)} \mu^+ \mu^-$ amplitude is
\be\label{eq:beta}
\beta_{SM} + \beta_{NP} \equiv -\frac{4G_F}{\sqrt{2}}V_{tb}^*V_{ts}
\frac{\alpha_{EM}(m_b)}{4\pi}  C_9^e + \frac{G}{2}\,
U^{d*}_{L33}U^d_{L32}|U^\ell_{L32}|^2 = -\frac{4G_F}{\sqrt{2}}V_{tb}^*
V_{ts}\frac{\alpha_{EM}(m_b)}{4\pi} C_9^\mu\,.
\ee
Note that the NP contribution to $C_9^e$ is negligible in our model. The LHCb
result for $R_K$ in Eq.~(\ref{eq:RK}) yields the useful ratio
\be\label{eq:rhoNP}
\rho_{NP} =\frac{\beta_{NP}}{\beta_{SM} + \beta_{NP}} = -0.159^{+
  0.069}_{-0.070}\,.
\ee
Then the branching ratio for $B^+ \to K^+\mu^\pm e^\mp$ (summed over lepton
charges) is given by
\be\label{eq:BKmue}
\CB(B^+ \to K^+\mu^\pm e^\mp) \cong 2\rho_{NP}^2
\left\vert\frac{U^\ell_{L31}}{U^\ell_{L32}}\right\vert^2 \CB(B^+ \to K^+
\mu^+\mu^-) = 
\left(2.16^{+2.54}_{-1.50}\right)\,\left\vert\frac{U^\ell_{L31}}
{U^\ell_{L32}}\right\vert^2 \times 10^{-8}\,,
\ee
where we used $\CB(B^+ \to K^+ \mu^+\mu^-) = (4.29\pm 0.22)\times 10^{-7}$ for
the branching ratio integrated over $q^2$~\cite{Aaij:2014pli}. The current
limit, $\CB(B^+ \to K^+\mu^\pm e^\mp) < 9.1\times
10^{-8}$~\cite{Agashe:2014kda}, gives the weak bound
\be\label{eq:BKmuelim}
|U^\ell_{L31}/U^\ell_{L32}| \simle 3.7\,.
\ee

Because the primary interaction $\CH_{NP}$ is in the third generation, the
decay $B^+ \to K^+ \mu^\pm \tau^\mp$ may be more interesting:
\be\label{eq:BKmutau}
\CB(B^+ \to K^+\mu^\pm \tau^\mp) \cong 2\rho_{NP}^2
\left\vert\frac{U^\ell_{L33}}{U^\ell_{L32}}\right\vert^2 \CB(B^+ \to
K^+\mu^+\mu^-)\,. 
\ee
The current limit, $\CB(B^+ \to K^+\mu^\pm \tau^\mp) < 4.8\times
10^{-5}$~\cite{Agashe:2014kda}, gives
\be\label{BKmutaulim}
|U^\ell_{L33}/U^\ell_{L32}| \simle 85 \,.
\ee
This is a crude estimate. It is performed by keeping only the terms
proportional to $|C_{9,10}|^2$ in the decay rate. It also neglects the
difference in the $q^2$ range and the phase space for the $\mu\mu$ and
$\mu\tau$ modes. These approximations affect $B \to K \tau^+\tau^-$ even
more. For this mode, only the weak limit $\CB(B \to K\tau^+\tau^-) <
3.3\times 10^{-3}$ has been set~\cite{Flood:2010zz}.

The $B_s$ decays to a pair of oppositely-charged leptons provide an interesting
correlation with $B \to K \ellp\ellm$. The only observed mode is
\be\label{eq:Bsmumu}
\CB(B_s \to \mu^+\mu^-)_{exp} = (2.8^{+0.7}_{-0.6})\times 10^{-9} =
(0.77 \pm 0.20)\times \CB(B_s \to \mu^+\mu^-)_{SM}\,,
\ee
where the experimental value is an average of LHCb and CMS measurements with
full Run~I statistics~\cite{Archilli_CKM2014}, while the SM value is $\CB(B_s
\to \mu^+\mu^-)_{SM} = (3.65\pm 0.23)\times
10^{-9}$~\cite{Bobeth:2013uxa}. The measurement is consistent with the SM
prediction, but also with $\CH_{NP}$ in Eq.~(\ref{eq:bsmmNP}) and the value
of $\rho_{NP}$ in Eq.~(\ref{eq:rhoNP}). Thus, our model implies the triple
correlation
\be\label{eq:reln}
R_K \cong \frac{\CB(B^+ \to K^+\mu^+\mu^-)_{exp}}{\CB(B^+ \to
  K^+\mu^+\mu^-)_{SM}}
\cong \frac{\CB(B_s \to \mu^+\mu^-)_{exp}}{\CB(B_s \to \mu^+\mu^-)_{SM}}\,.
\ee
This relation identifies the numerical factor on the right of
Eq.~(\ref{eq:Bsmumu}) with Eq.~(\ref{eq:RK}) and stresses the importance of a
more accurate measurement of $\CB(B_s \to \mu^+\mu^-)$. The only reported LFV
limit, $B(B_s \to \mu^\pm e^\mp) < 1.1 \times 10^{-8}$~\cite{Aaij:2013cby},
gives $|U^\ell_{L31}/U^\ell_{L32}| \simle 35$, an order of magnitude weaker
than the bound 3.7 from $B^+ \to K^+ \mu e$.  We hope that searches for this
mode and for $B_s \to \mu\tau$ in LHC Run~II can provide much improved
limits.\footnote{We have not examined the interesting possibility that phases
  in the off-diagonal $U^{d,\ell}$ matrix elements may induce new
  $CP$-violating effects, especially in $B_s$ decays.}

Measurements exist for $\CB(B^0 \to K^0 \ellp\ellm)$ ($\ell = e,\mu$) and
$\CB(B_s \to \phi\mu^+\mu^-)$. These are comparable to but less precise than
$\CB(B^+\to K^+\ellp\ellm)$. Similarly, limits on $B^0 \to \ellp\ellm$ and
the LFV decays $B^0 \to K^{(*)0} \mu e$ and $\ell^{\pm}\ell^{'\mp}$,
including $\tau$-modes, give no more stringent limits than those for $B^+$
decays. The amplitudes for $B^0 \to \ellp\ellm$ and $\ell^\pm\ell^{'\mp}$ are
suppressed by $V_{td}/V_{ts}$ and $U^d_{L31}/U^d_{L32}$ relative to the
corresponding $B_s$ decays.\footnote{Our interaction $\CH_{NP}$, as well as
  those mentioned in footnote~1, will induce rare LFV decays for kaons. Such
  a study is worthwhile because of the considerable interest world-wide in
  experiments with intense beams, but it is beyond the scope of this paper.}

Finally, the operator
\be\label{eq:MECO}
\CH_{NP}(\mu + d \to d + e) = G\, |U^d_{L31}|^2 \, \bar d_L \gamma^\lambda
d_L \left(U^{\ell *}_{L31} U^\ell_{L32} \, \bar e_L \gamma_\lambda \mu_L +
{\rm H.c.}\right)
\ee
induces $\mu \to e$ conversion in nuclei.\footnote{Our model Hamiltonian
  $\CH_{NP}$ can also induce $\mu \to e \gamma$. Simply closing up the quark
  line and emitting a photon from it does not produce the desired operator,
  $\propto e\,G \,m_\mu\, \bar e\sigma^{\lambda\nu} \mu F_{\lambda\nu}$.
  Virtual exchanges of electroweak or Higgs bosons and/or a mass insertion
  must occur between the quark and lepton lines to change the muon chirality.
  We expect that such an operator is too weak to give an interesting limit.}
The limit on the strength of this operator from conversion in Titanium is
(see Table~3.6 in Ref.~\cite{Raidal:2008jk}):
\be\label{eq:MECOlimita}
G\,|U^d_{L31}|^2 |U^{\ell *}_{L32} U^\ell_{L31}| < \frac{4G_F}{\sqrt{2}}
\times (8.5 \times 10^{-7})\,.
\ee
Using
\be\label{eq:betanpsm}
\left\vert\frac{\beta_{NP}}{\beta_{SM}}\right\vert = 
\left\vert\frac{\pi \,G \,U^{d*}_{L33}U^d_{L32} |U^\ell_{L32}|^2} {\sqrt{2}
    G_F\,\alpha_{EM}\,V^*_{tb} V_{ts} \,C_9^e}\right\vert \simeq 0.14\,,
\ee
together with $|V_{tb}| \simeq 1$, $|U^d_{L32}| \simeq |V_{ts}| \cong 0.043$,
$|U^d_{L31}| \simeq |V_{td}| \cong 0.0084$~\cite{Agashe:2014kda}, and
$\alpha_{EM}(m_b) = 1/133$, we obtain the weak limit
\be\label{eq:MECOlimitb}
\left\vert\frac{U^\ell_{L31}}{U^\ell_{L32}}\right\vert < \frac{75}{|
C_9^e|} \cong 18\,,
\ee
where $|C^e_9| \cong 4.07$~\cite{Buchalla:1995vs}. The Mu2e experiment at
Fermilab is designed to be sensitive to an interaction strength 100 times
smaller than in Eq.~(\ref{eq:MECOlimita}). Under our assumptions, Mu2e would
then be able to probe $|U^\ell_{L31}/U^\ell_{L32}| \simge 0.2$.

Summing up: The interesting new results on $B\to K^{(*)} \mu^+\mu^-$ and $R_K$
from LHCb, if correct, tell us that there are lepton number non-universal
interactions. Therefore, there must also be lepton flavor violating
interactions, and there is no known reason these should be very much weaker
than the non-universal ones. Limits from searches for $B \to K
\ell^+\ell^{\prime -}$ and $B_s \to \ell^+\ell^{\prime -}$ are not far above
interesting ranges for LFV mixing-angle parameters. LHCb's results make
searches for these and other rare processes well worth pursuing.

\section*{Acknowledgments} 

We are grateful for stimulating conversations with Adam Martin and Edwige
Tournefier. We have also benefited from discussions with Wolfgang
Altmannshofer, Francesco Dettori, Federico Mescia, Jim Miller, Marco Nardecchia,
Chris Quigg, David Straub, Danny van~Dyk and Roman Zwicky. SLG's research is
supported in part by the U.S.~Department of Energy under Grant
No.~DE-SC0010025. KL gratefully acknowledges support of this project by the
Labex ENIGMASS during 2014. He thanks Laboratoire d'Annecy-le-Vieux de
Physique Th\'eorique (LAPTh) for its hospitality and the CERN Theory Group
for support and hospitality.  KL's research is supported in part by the
U.S.~Department of Energy under Grant No,~DE-SC0010106.


\bibliography{LFV}
\bibliographystyle{utcaps}
\end{document}